\documentclass[prb,twocolumn,
floatfix,showpacs]{revtex4}
\usepackage{graphicx}
\usepackage{times}
\newcommand{\be}{\begin{equation}}
\newcommand{\ee}{\end{equation}}
\newcommand{\bea}{\begin{eqnarray}}
\newcommand{\eea}{\end{eqnarray}}
\newcommand{\bm}[1]{\mbox{\boldmath$#1$}}
\newcommand{\ra}{\rangle}
\newcommand{\la}{\langle}
\begin{document}
\title{Universality in the diffusion of knots 
}

\author{Naoko Kanaeda}
\email{kanaeda@degway.phys.ocha.ac.jp}
\author{Tetsuo Deguchi}
\email{deguchi@phys.ocha.ac.jp}
\affiliation{Department of Physics, 
Graduate School of Humanities and Sciences, 
Ochanomizu University, 2-1-1
Ohtsuka, Bunkyo-ku, Tokyo 112-8610, Japan}

\date{\today}

\begin{abstract}
We have evaluated a universal ratio between diffusion constants 
of the ring polymer with a given knot $K$ 
and a linear polymer with the same molecular weight 
in solution through the Brownian dynamics under hydrodynamic interaction.  
The ratio is found to be constant 
with respect to the number of monomers, $N$, and hence 
the estimate at some $N$  
should be valid practically over a wide range of $N$ 
for various polymer models. 
Interestingly, the ratio is determined by  
the average crossing number ($N_{AC}$) 
of an ideal conformation of knotted curve $K$,  
i.e. that of the ideal knot.  The $N_{AC}$ of ideal knots 
should therefore be fundamental in the dynamics of knots.   
\end{abstract}

\pacs{82.35.Lr,05.40.Fb,05.20.-y}

%
%

%

\maketitle

\section{Introduction} 
Novel knotted structures of polymers have recently been found 
in various research fields such as DNA, proteins and synthetic polymers. 
\cite{Cieplak,Rabin,Baiesi,Ercolini}  
The topology of a ring polymer is  
conserved under thermal fluctuations in solution 
and represented by a knot. 
\cite{Casassa,Vologodskii,Roovers,Semlyen,tenBrinke} 
Topological constraints may lead to nontrivial statistical mechanical 
and dynamical properties of ring polymers \cite{Vologodskii,Muthukumar,Quake94,Univ97,GrosbergPRL00,Lai02,Shima02,Akos03,Zonta05,Orlandini07,Whitt07}.

Recent progress in experiments of ring polymers should be quite remarkable.  
Diffusion constants of linear, relaxed circular and supercoiled 
DNAs have been measured quite accurately. \cite{PNAS06} 
Here the DNA double lelices are unknotted. 
Furthermore, hydrodynamic radius of circular DNA 
has also been measured. \cite{Araki06} 
Ring polymers of large molecular weights 
are synthesized not only quite effectively \cite{Bielawski}  
but also with small dispersions and of high purity. \cite{Takano05,Takano07} 
Circular DNAs with various knot types are derived, 
and they are separated into knotted species 
by gel electrophoresis. \cite{ACN} We should remark that  
synthetic ring polymers with nontrivial 
knots have not been synthesized and separated experimentally, yet.  
However, it is highly expected that ring polymers of  nontrivial knot types 
should be synthesized and their topological effects 
will be confirmed experimentally in near future.

In the paper we discuss diffusion constant $D_K$  
of a ring polymer with fixed topology $K$ in good solution 
for various knot types. We evaluate it numerically 
 via the Brownian dynamics with hydrodynamic interaction 
in which bond crossing is effectively prohibited through the 
finite extensible non-linear elongational (FENE) potential. \cite{IOP} 
We evaluate diffusion constant $D_L$  
of a linear polymer with the same molecular weight, 
and derive  ratio $D_K/D_L$. The ratio should corresponds to a universal 
amplitude ratio of critical phenomena and 
play a significant role in the dynamics of knotted ring polymers. 
According to the renormalization group arguments, ratio $D_K/D_L$ 
should be universal if the number of monomers, $N$, is large enough. 
\cite{Oono,Oono-Kohmoto,Schaub}

The ratio $D_K/D_L$ may have some experimental applications. 
Ring polymers of different knot types can be separated 
experimentally with respect to their topologies by   
making use of the difference among the sedimentation coefficients, 
which can be calculated from the diffusion constants. 
\cite{Zimm1966}
Here we remark that 
the diffusion constant of a ring polymer under no topological constraint 
, $D_{R}$, and that of the corresponding linear polymer 
has been numerically evaluated, and  the ratio $C=D_{R}/D_L$  
has been studied. \cite{Bernal90,Bernal91,IOP}

Through simulation we find that ratio $D_K/D_L$ is approximately constant 
with respect to $N$ for various knots. Thus, if we evaluate 
ratio $D_K/D_L$ at some value of $N$,  it is practically valid 
for other values of $N$.
We can therefore predict the diffusion constant 
$D_K$ of a polymer model at some value of $N$, 
multiplying the ratio $D_K/D_L$ by the estimate of $D_L$.  
Here we remark that the value of $D_L$ 
 may depend on the number $N$ and on some details of the model. 
\cite{Dunweg,Liu}

Furthermore, we show numerically that ratio $D_K/D_L$ 
is a linear function of the average crossing number ($N_{AC}$) 
of the ideal knot of $K$,  an ideal  configuration of knotted curve $K$,
which will be defined shortly. Since the ratio $D_K/D_L$  
is almost independent of $N$, it follows that the linear fitting formula 
should be valid practically in a wide range of finite values of $N$.  
Thus, the ideal knot of a knotted curve $K$ should play a fundamental role 
in the dynamics of finite-size knotted ring polymers in solution.

Let us introduce the ideal knot, briefly.  For a given knot $K$ 
it is given by the trajectory 
that allows maximal radial expansion 
of a virtual tube of uniform diameter centered around the 
axial trajectory of the knot $K$. \cite{Grosberg96,Katritch} 
We define the $N_{AC}$ of a knotted curve as follows: 
We take its projection onto a plane, and   
enumerate the number of crossings in the knot diagram on the plane.    
Then, we consider a large number of projections onto planes 
whose normal vectors are 
uniformly distributed on the sphere of unit radius,  
and take the average of the crossing number ($N_{AC}$)  
over all the normal directions.  

The paper consists of the followig: In section II, the simulation 
method is explained. In section III, we present the 
 estimates of the diffusion constant of a ring polymer in solution 
of knot type $K$ for various knot types.  Then, we show numerically that 
the graph of $D_K/D_0$ is almost independent of $N$, 
and also that ratio $D_K/D_L$ is fitted by a linear function of $N_{AC}$ of 
the ideal knot of $K$. We also discuss the simulation result 
in terms of the ratio of equivalent radii, \cite{Ortega}  $a_G/a_T$, 
which corresponds to the universal ratio of the radius of 
 gyration to the hydrodynamic radius. \cite{Dunweg} 
We shall define the equivalent radii explicitly in section III. 
Finally, we give conclusion in section IV.  

Throughout the paper, we employ the symbols of knots 
following Rolfsen's textbook \cite{Rolfsen}, as shown in Figure 1.

\section{Simulation method}
The ring polymer is modeled as a cyclic bead-and-spring chain 
with $N$ beads connected by $N$ FENE springs with force given by   
\be 
{\bm F}({\bm r})=- H {\bm r}/(1-r^2/r^2_{max})
\ee
where $r=|{\bm r}|$. We denote by $b$ the unit of distance, which gives  
the average distance between neighboring monomers approximately.   
We set constants $H$ and $r_{max}$ by $H=30 k_B T/b^2$ and $r_{max}=1.3 b$. 
We assume the Lennard-Jones (LJ) potential by 
\be
V(r_{ij}) = 4 \epsilon_{\rm LJ} \{ \left({\sigma_{\rm LJ}}/{r_{ij}} 
\right)^{12} - \left({\sigma_{\rm LJ}}/{r_{ij}} \right)^{6} \} \, . 
\ee 
Here $r_{ij}$ is the distance of beads $i$ and $j$, and $\epsilon_{\rm LJ}$  
and $\sigma_{\rm LJ}$ denote the minimum energy and the zero energy distance, 
respectively. \cite{Cifre} 
We set the Lennard-Jones parameters 
as $\sigma_{LJ}=0.8 b$ and $\epsilon_{LJ}=0.1 k_{\rm B} T$ 
so that they give good solvent conditions. \cite{Rey}    
Here $k_B$ denotes the Boltzmann constant. 

We employ the predictor-corrector version \cite{Iniesta} 
of the Ermak-McCammon algorithm 
for generating time evolution of 
a ring polymer in solution. 
The hydrodynamic interaction is taken into account 
through the Rotne-Prager-Yamakawa tensor 
\cite{Rotne,Yamakawa,Cifre} 
where the bead friction is given by $\zeta=6 \pi \eta_s a$ 
with the bead radius $a=0.257b$ and a dimensionless 
hydrodynamic interaction parameter 
$h^{*}=(\zeta/6 \pi \eta_s) \sqrt{H/\pi k_B T}= 0.25$.  

In the present simulation, physical quantities are given in dimensionless 
units such as in Ref. \cite{Cifre}.  
We divide length by $b$, energy by $k_B T$ and time 
by $\zeta b^2/k_{\rm B} T$. 
Let us indicate dimensionless quantities by an asterisk as superscript.  
We have $H^{*} = 30$, $r^{*}_{max}=1.3$. 
We take the simulation time step $\Delta t^{*}= 10 ^{-4}$. 

We have set the FENE potential so that 
the topology of the ring polymer should be effectively conserved. \cite{IOP} 
However, bond crossing may occur with a very small probability.  
Calculating knot invariants, we have confirmed 
that the fraction of nontrivial knots is very small.  
If the initial knot type is the trivial knot 
it is given by $10^{-8} \sim 10^{-7}$, and 
if the initial knot type is a nontrivial knot, 
it is given by approximately $10^{-7}$.  

\section{Simulation results}
We define the diffusion constant of a polymer by  
\be 
D =  \lim_{t \rightarrow \infty} {\frac 1 {6t}} 
\langle (\vec{r}_G(t)-\vec{r}_G(0))^2  \rangle \,. 
\label{eq:dfD}
\ee
Here ${\vec r}_G(t)$ denotes the position vector of the center of mass 
of the polymer at time $t$. 
Making use of (\ref{eq:dfD}) 
we have evaluated diffusion constants $D_L$ and $D_K$.

\begin{table} \begin{center} \begin{tabular}{ccccc}
knot type &$D$&$\langle R_G^2 \rangle$\\ \hline
linear &$0.12038\pm 0.00085$   &$9.33029\pm0.03219$ \\
0   &$0.13059\pm 0.00089$   &$5.26539\pm0.00913$ \\
$3_1$     &$0.14530\pm 0.00079$   &$3.21052\pm 0.00505$ \\
$4_1$     &$0.14876\pm 0.00074$   &$2.78817\pm 0.00160$ \\
$5_1$     &$0.15277\pm 0.00085$   &$2.72300\pm 0.00261$ \\
$5_2$     &$0.15640\pm 0.00078$   &$2.61427\pm 0.00132$ \\
$6_1$     &$0.15927\pm 0.00078$   &$2.47449\pm 0.00137$ \\
$6_2$     &$0.15902\pm 0.00095$   &$2.37272\pm 0.00126$ \\
$7_1$     &$0.16416\pm 0.00083$   &$2.47162\pm 0.00109$ \\
\hline \end{tabular}
\caption{Estimates of diffusion constants $D_L$ and $D_K$, and 
the mean square radius of gyration $\langle R^2_G \rangle$ and   
 for a linear polymer of $N=45$ and ring polymers 
of $N=45$ with various knot types. }
\end{center} \end{table}

The estimates of diffusion constants  
$D_L$ and $D_K$ at $N=45$ are listed in Table 1 
together with those of the mean square radius 
of gyration $\langle R_G^2 \rangle$. 
The data of $D_L$ and $D_K$ are plotted against $N$ in Figure 2. 
The fitting curves to them are given by $D=a N^{-\nu}(1+b N^{-\Delta})$. 
Here the errors of the diffusion constants are as small as $10^{-4}$.

Ratios $D_K/D_L$ should correspond to universal amplitude ratios 
in critical phenomena. Numerically we find  
that ratio $D_{K_1}/D_{K_2}$ of two different knots $K_1$ and $K_2$ 
is almost constant with respect to $N$, at least in the range investigated . 
For instance, the graph of ratio $D_{3_1}/D_0$ versus $N$ 
and that of ratio $D_{4_1}/D_0$ versus $N$ 
for the data are almost flat, as shown in Figures 3 and 4, 
respectively.  
Here $0$, $3_1$ and $4_1$ denote the trivial,  the trefoil and 
the figure-eight knot, respectively, as shown in Figure 1.   
The numerical values of $D_{3_1}/D_0$ are given from 1.14 to 1.17 
in Figure 3, 
and those of  $D_{4_1}/D_0$ are given from 1.14 to 1.21 in Figure 4. 
Thus, the estimate of $D_{K}/D_0$ evaluated at a value of $N$, 
say $N=45$, for some knot $K$ should also be valid 
at other finite values of $N$,  
since it is almost independent of $N$.

For the diffusion constant of a ring polymer, $D_{R}$, 
the ratio $D_R/D_L$ should 
correspond to a universal amplitude ratio and should be universal 
if $N$ is large enough. \cite{Oono,Oono-Kohmoto,Schaub} 
For the diffusion constant $D_R$,   
there is no topological constraint  
in the ring polymer model and $D_{R}$ does not mean $D_K$ of a knot $K$. 
\cite{Bernal90,Bernal91,IOP} 
In the previous simulation  \cite{IOP} 
it has been found that ratio $D_{0}/D_L$ is given by about 1.1 
for the present polymer model and almost 
independent of $N$ within the range investigated.

 From the numerical observations and the RG arguments,  
we have two conjectures: (A) $D_{0}/D_L$ should be given 
by 1.1 for some wide range of finite values of $N$ and 
also in the large $N$ limit; (B) ratio $D_{K}/D_0$ 
for a nontrivial knot $K$ should remain almost the same value 
in a wide range of finite values of $N$, 
i.e. the $N$-dependence should be very small.

Quite interestingly we find that  ratio $D_K/D_L$ can be approximated 
by a linear function of the average crossing number ($N_{AC}$) of ideal knots, 
i.e. the ideal representations of the corresponding knots.    
In Figure 5 
simulation data of $D_K/D_L$ 
are plotted against $N_{AC}$ of ideal knots.  
We find that the data points are fitted well 
by the following empirical formula:  
\be 
D_K/D_L= a + b \, {N_{AC}} . \label{eq:ACN}
\ee
Here, the estimates of $a$ and $b$ 
are given in the caption of  Figure 5.  
Thus, the diffusion constant $D_K$ of a knot 
$K$ can be estimated in terms of the $N_{AC}$ 
of the ideal knot of $K$.

Let us discuss the $\chi^2$ values. 
We have $\chi^2=2$ for the fitting line of Figure 5, which 
is for the data of $N=45$. For the data of $N=36$ 
we have a good fitting line with $\chi^2=3$.  
The estimates  
of $a$ and $b$ for $N=36$ are similar to those for $N=45$. 
Thus, we may conclude that the graph of $D_K/D_L$ versus $N_{AC}$ 
is fitted by a linear line.

For a finite value of $N$, we can estimate 
the diffusion constant $D_K$ of a knot $K$ 
through formula (\ref{eq:ACN}) 
by the $N_{AC}$ of the ideal knot of $K$.
Here we have assumed that coefficients $a$ and $b$ of (\ref{eq:ACN}) 
are independent of $N$ since the graphs of $D_K/D_0$ 
and $D_0/D_L$ are almost flat with respect to $N$.     
In fact, there is almost no numerical support 
for suggesting a possible $N$-dependence of $a$ and $b$, directly.

We thus summarize the simulation results so far as follows:    
ratio $D_K/D_0$ for a knot $K$ should be almost constant 
with respect to $N$ in a wide range of $N$ 
and can be expressed by the linear function of $N_{AC}$ of ideal knots.  
Eq. (\ref{eq:ACN}) should be useful in separating synthetic 
ring polymers into various knotted species by making use of 
the difference among sedimentation coefficients. 

Ideal knots should play a fundamental role in the dynamics 
of knotted ring polymers in solution. 
In fact, we have shown it for the diffusion constants. 
In experiments of gel  electrophoresis drift velocities of different knots  
formed on the same DNA molecules were shown 
to be simply related to the $N_{AC}$ of ideal knots. \cite{ACN}  
The two independent results suggest the importance 
of the $N_{AC}$ of ideal knots in the dynamics of knotted ring polymers,  
although the physical situations are different.

Let us now discuss the simulation results from the viewpoint of  
equivalent radii. \cite{Ortega} 
The equivalent radius for any solution property is the radius 
of a spherical particle having the same value of solution property as 
that of the macromolecule under consideration. 
The ratio of equivalent radii should  be universal, and it 
should play a similar role as 
the universal amplitude ratio such as 
the ratio of diffusion constants. \cite{Dunweg} 
We define equivalent radii $a_G$ and $a_T$ explicitly by 
\bea 
a_{G} & = & \sqrt{ {\frac 5 3} \langle R^2_G \rangle } \, , \\
a_{T} & = & {\frac {k_{\rm B} T} {6 \pi \eta_s D }} \, .   
\eea 
Here $a_G$ and $a_T$ corresponds to the radius of gyration 
$R_G=\sqrt{\la R_G^2 \ra }$ and the translational friction 
coefficient $D$, resepctiveley. 
The ratio $a_G/a_T$ 
corresponds to the ratio of the radius of 
 gyration to the hydrodynamic radius,  
and should be universal.

The numerical estimates of $a_G/a_T$ for $N=45$ 
for the present simulation are listed in Table 2 for 
linear and ring polymers with various knot types. 
In Figure 6, the ratio $a_G/a_T$ is plotted against the number of segments, 
$N$,  for  linear and ring polymers with various knot types. 
Interestingly, the graphs show a weak $N$-depndence.   
They are fitted by a function $a_G/a_T = a(1 - b N^{-c})$ 
with parameters $a$, $b$ and $c$ being positive. 
It suggests that the graphs become constant with respect to $N$  
if $N$ is large enough. We thus expect that the ratio $a_G/a_T$ 
in the large $N$ limit should be universal.

It is interesting to note in Figure 6 that the estimate of ratio $a_G/a_T$ 
in the large $N$ limit is distinct for the different topologies 
such as linear polymers and ring polymers of the trivial and 
trefoil knots. The ratio could thus be useful for 
detecting the knot type of a ring polymer in solution.

\begin{table} \begin{center} \begin{tabular}{ccc}
knot type & $a_G/a_T$\\ \hline
linear &$1.8471 \pm 0.00475$\\
0   & $1.5053 \pm 0.01869$\\
$3_1$     &$1.3079 \pm 0.01151$\\
$4_1$     &$1.20824 \pm 0.00055$\\
$5_1$     &$1.26636 \pm 0.00077$\\
$5_2$     &$1.27029 \pm 0.00052$\\
$6_1$     &$1.25865 \pm 0.00084$\\
$6_2$     &$1.23046 \pm 0.00058$\\
$7_1$     &$1.29643 \pm 0.00050$\\
\hline \end{tabular}
\caption{Estimates of ratio  
$a_G/a_T=\sqrt{5 \langle R_G^{2*} \rangle/3} \, 
{D^*}/a^{*}$  for a linear polymer of $N=45$ and ring polymers 
of $N=45$ with various knot types. Here, 
$\langle R_G^{2*} \rangle= \langle R_G^2 \rangle/b^2$, 
$D^{*}= 6 \pi \eta_s a D/k_B T$ and $a^{*}=a/b$. 
}
\end{center} \end{table}

%

\section{Conclusion}
We have evaluated universal ratios among 
the diffusion constants of knotted ring polymers 
in good solution for several knots, 
 where bond crossing is effectively prohibited 
 in the Brownian dynamics under hydrodynamic interaction. 
 The universal ratio of 
diffusion constants $D_K/D_L$ is almost constant with respect to 
the number of polymer segments, $N$. 
Moreover, it is found that the ratio $D_K/D_L$ is determined  
by the $N_{AC}$ of the ideal knot of $K$. 
Through the linear relation,  
we can estimate the diffusion constant of a given knot.

\section*{Acknowledgments}
The authors would like to thank Dr. A. Takano and 
Dr. K. Tsurusaki for valuable comments. 
The present study is partially supported 
by KAKENHI (Grant-in-Aid for Scientific Research) on Priority Area 
``Soft Matter Physics'' from the Ministry of Education, 
Culture, Sports, Science and Technology of Japan, 19031007. 
We drew the figures of a linear and knotted ring polymers using OCTA($\langle$ http://octa.jp $\rangle$).

\newpage 
\begin{figure}
  \begin{center}
    \begin{tabular}{ccc}
      \resizebox{40mm}{!}{\includegraphics{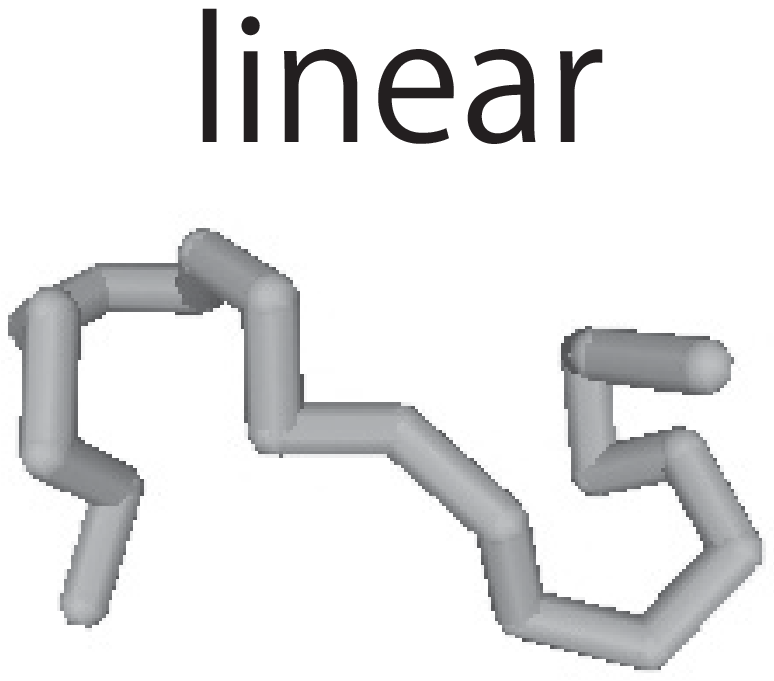}} &
      \resizebox{40mm}{!}{\includegraphics{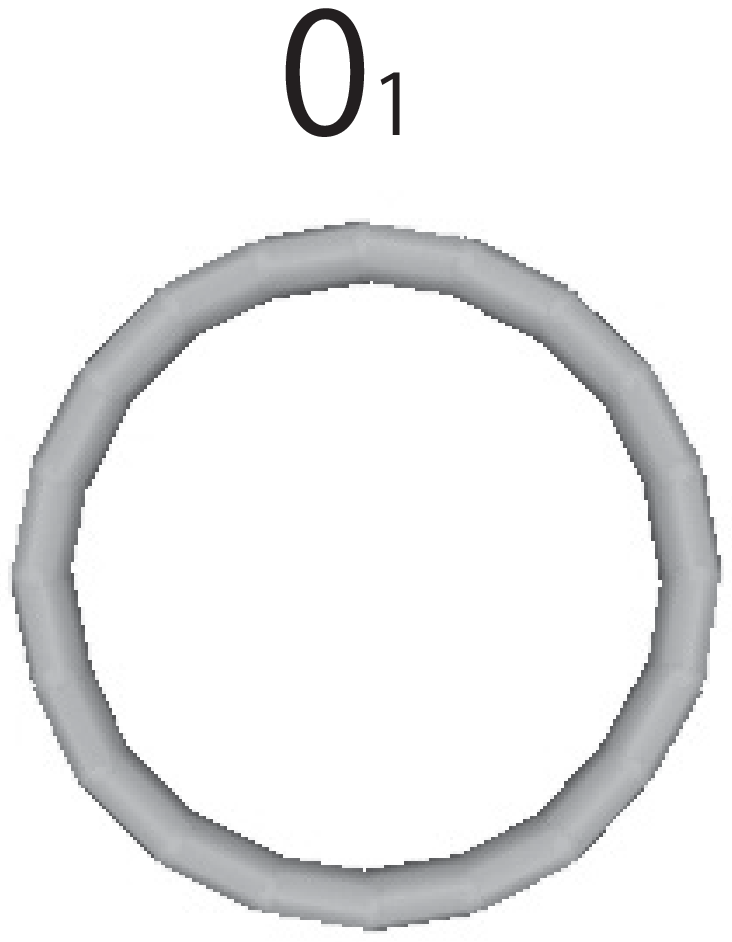}} &
      \resizebox{40mm}{!}{\includegraphics{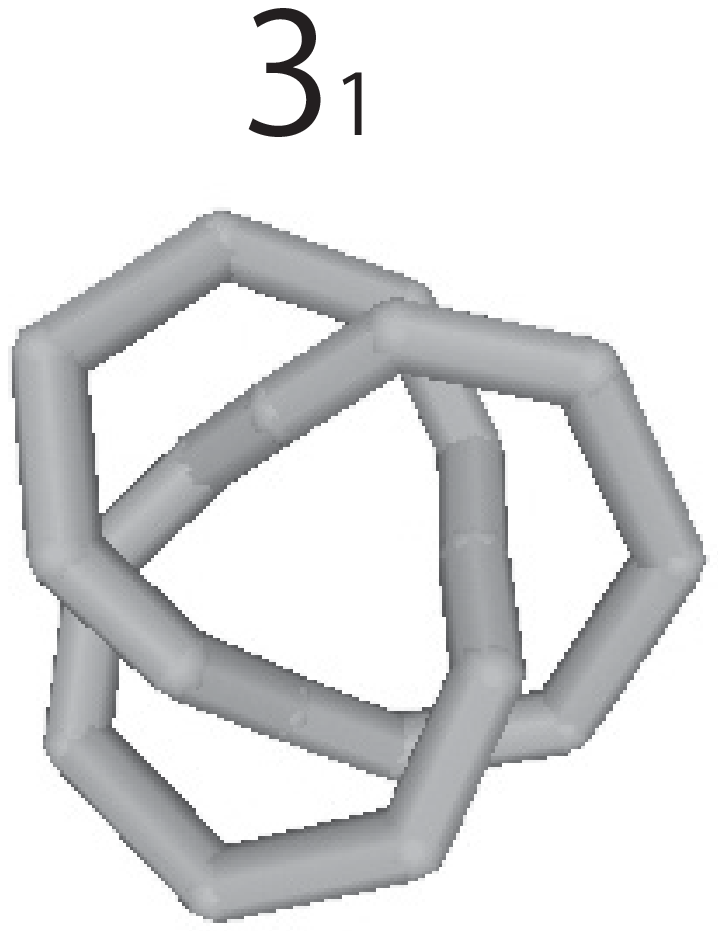}} \\
      \resizebox{40mm}{!}{\includegraphics{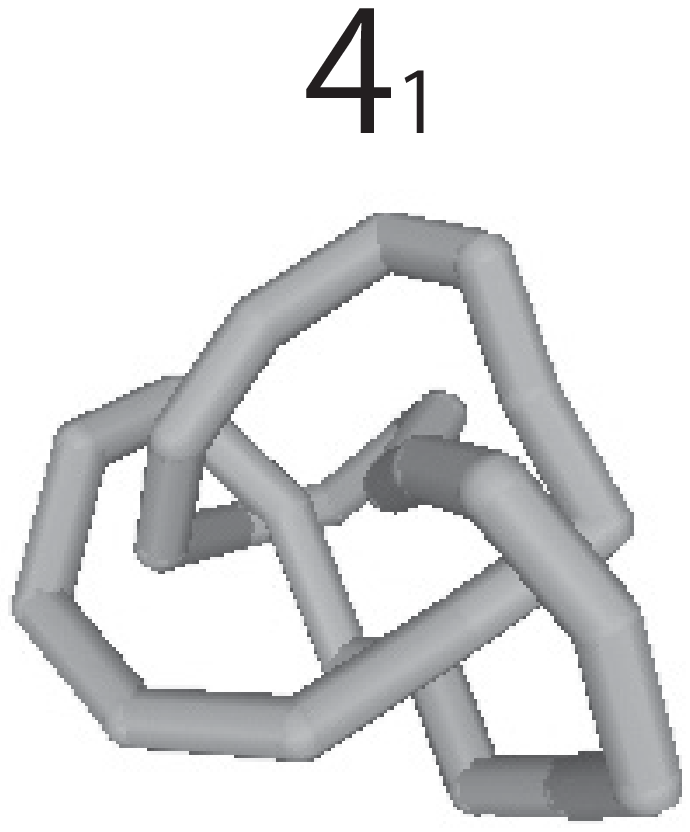}} &
      \resizebox{40mm}{!}{\includegraphics{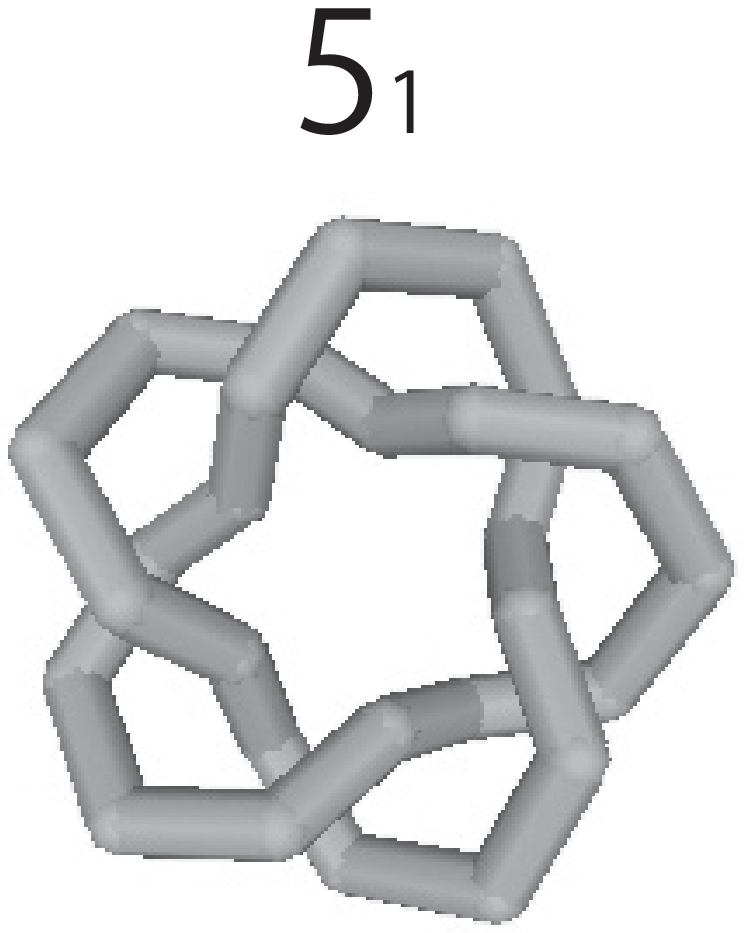}} &
      \resizebox{40mm}{!}{\includegraphics{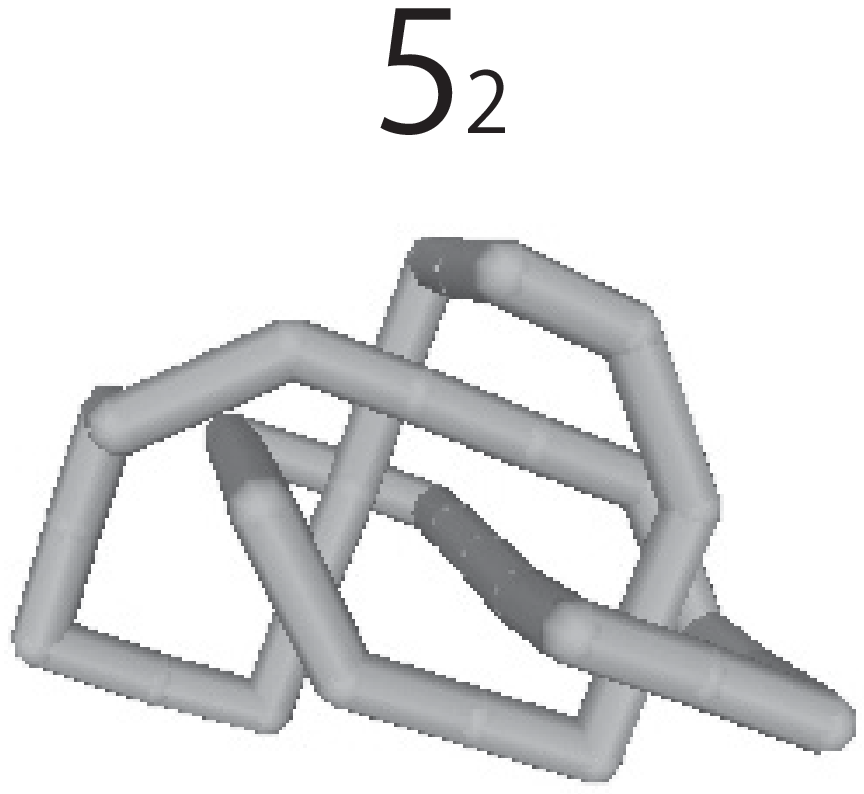}} \\
      \resizebox{40mm}{!}{\includegraphics{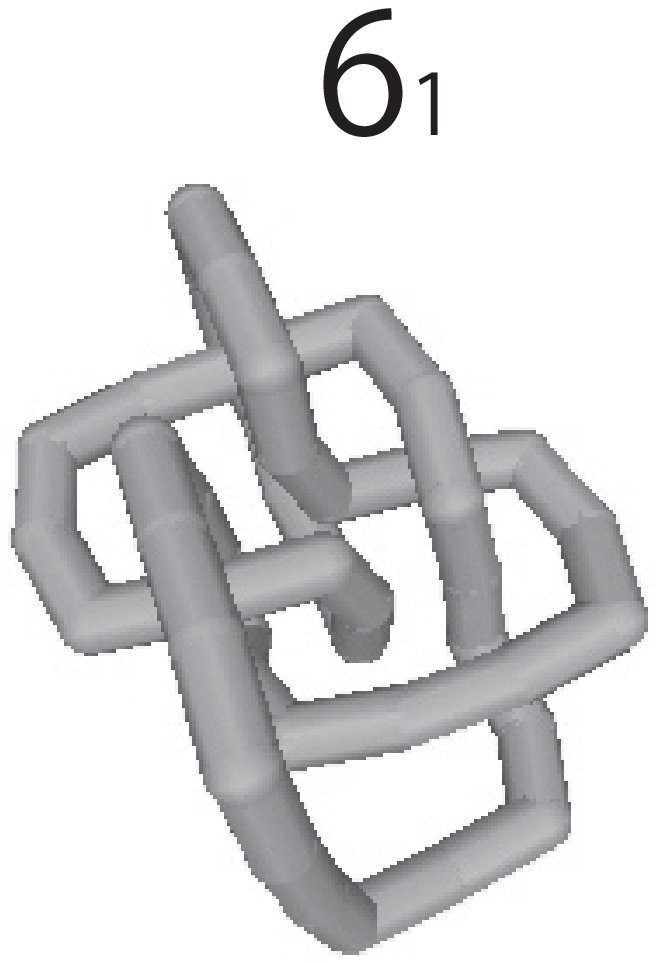}} &
      \resizebox{40mm}{!}{\includegraphics{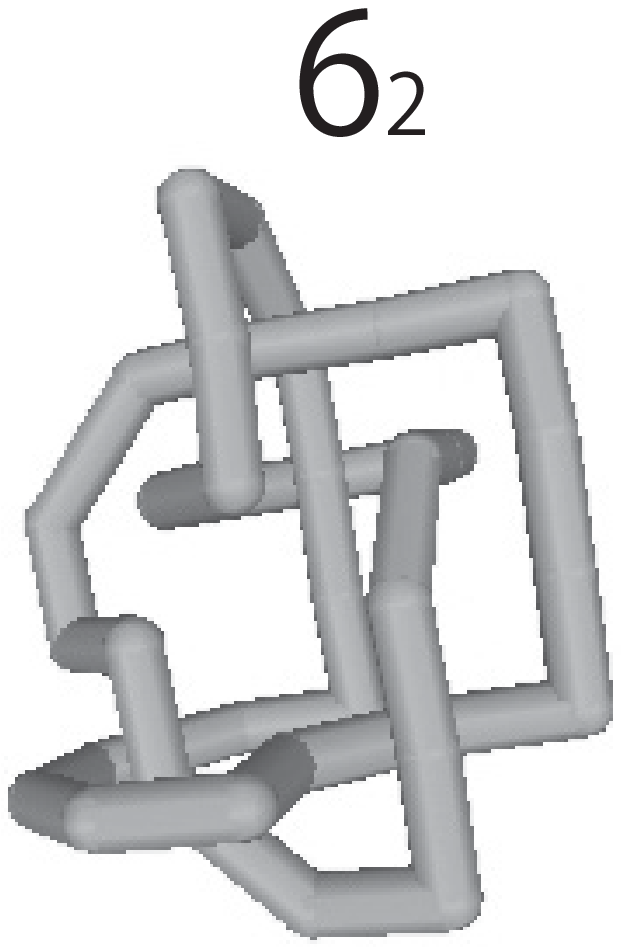}} &
      \resizebox{40mm}{!}{\includegraphics{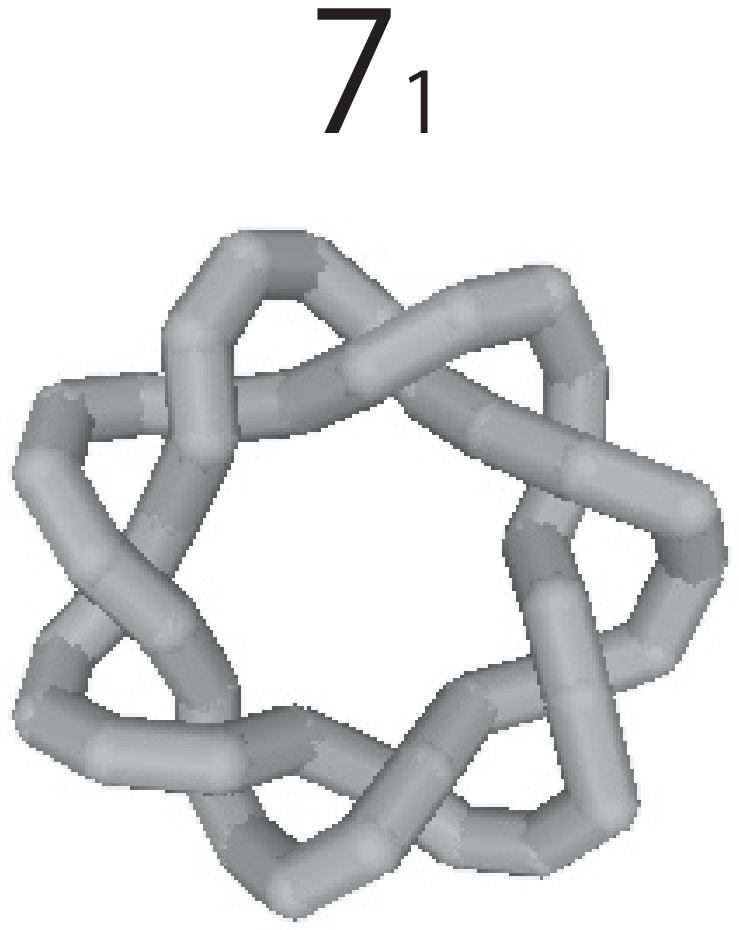}} \\
    \end{tabular}
    \caption{Figures of a linear polymer and knotted ring polymers 
with the symbols of knots given in Rolfsen's textbook. }
    \label{fig:polymers}
  \end{center}
\end{figure}

\begin{figure}[htb]
  \begin{center}
    \begin{tabular}{cc}
  \resizebox{120mm}{!}{\includegraphics{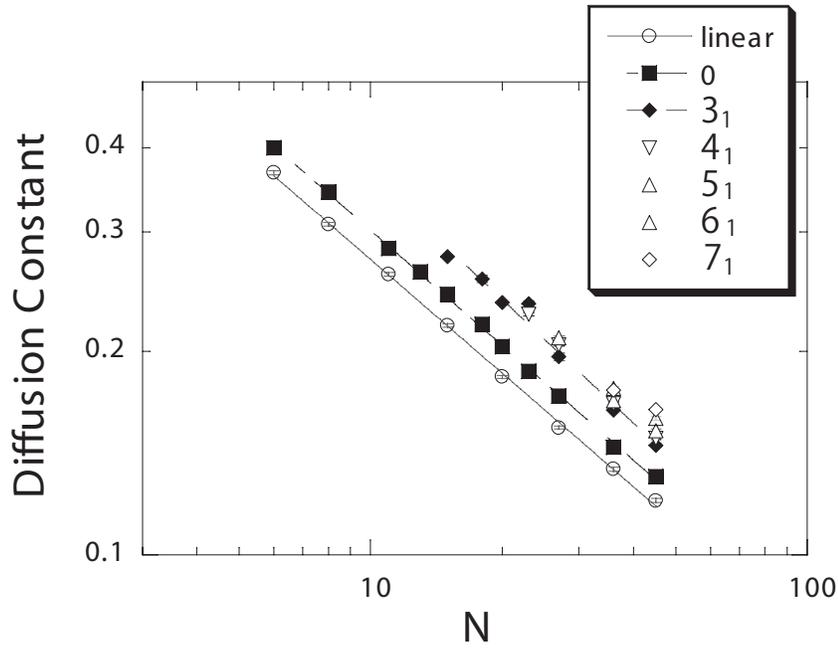}}
    \end{tabular}
    \label{fig:diff_const}
  \end{center}
\caption{Diffusion constants of linear and knotted ring chains 
with knots $0$, $3_1$, $4_1$, $5_1$, $6_1$ and $7_1$, versus 
$N$. 
 Fitted by $D = a N^{-\nu}(1 + b N^{-\Delta})$ with  
the following best estimates: 
For a linear chain, 
$a=0.90 \pm 0.23$, $\nu=0.53 \pm 0.06$, 
$b=0.51 \pm 0.93$, $\Delta=1.14 \pm 2.39$, $\chi^2=17$; 
for the trivial knot (0), 
 $a=1.03 \pm 1.11$, $\nu=0.55 \pm 0.18$, $b=0.14 \pm 0.78$, 
$\Delta= 0.60 \pm 6.09$, $\chi^2=28$; 
for  the trefoil knot ($3_1$) 
$a=1.00 \pm 3.87$, $\nu =0.52 \pm 0.67$, $b=1.18 \pm 1.12$,  
$\Delta=0.77 \pm 6.09$, $\chi^2=27$. }
\end{figure}

%
%

\begin{figure}
  \begin{center}
    \begin{tabular}{cc}
  \resizebox{120mm}{!}{\includegraphics{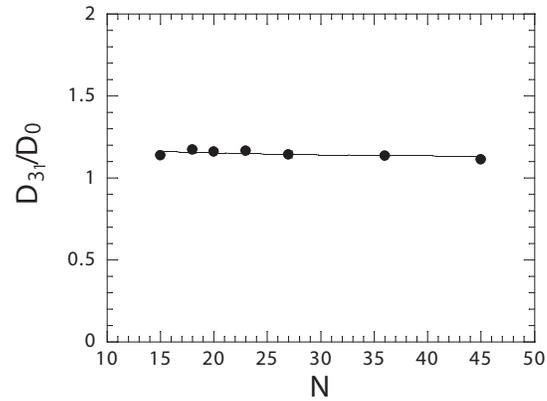}}
    \end{tabular}
    \label{fig:N-independence1}
  \end{center}
\caption{Ratio $D_{3_1}/D_{0}$ of diffusion constants 
for the trefoil knot ($3_1$) and the trivial knot ($0$) 
versus the number of segments $N$. Fitting curve is given by 
$D_{3_1}/D_{0}=a(1+b N^{-c})$, where 
$a=1.07 \pm 0.64$, $b=0.25 \pm 0.39$, 
and $c=0.39 \pm 3.29$ with $\chi^2=6$. }
\end{figure}


\begin{figure}
  \begin{center}
    \begin{tabular}{cc}
  \resizebox{120mm}{!}{\includegraphics{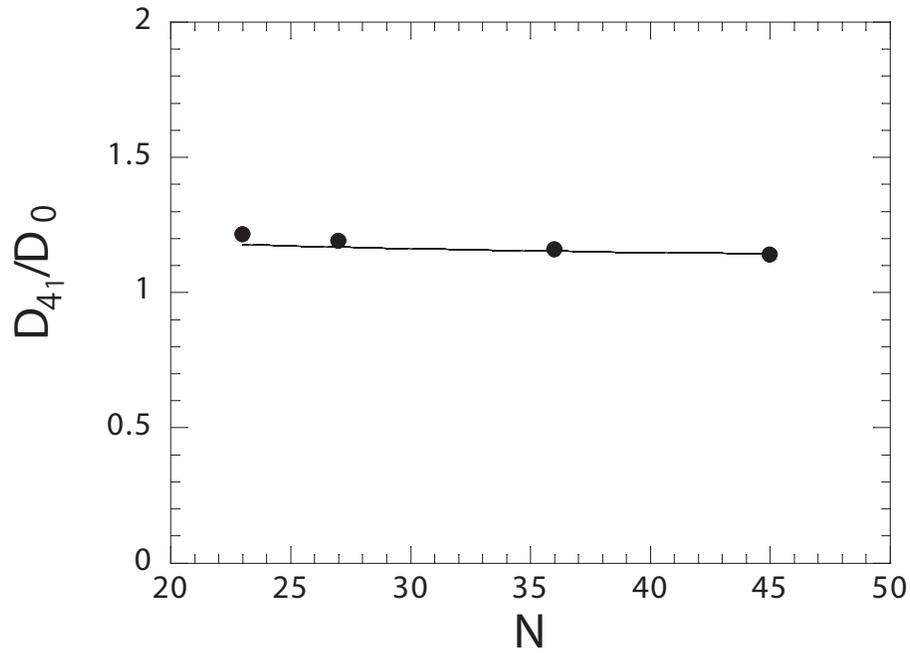}}
    \end{tabular}
    \label{fig:N-independence2}
  \end{center}
\caption{Ratio $D_{4_1}/D_{0}$ of diffusion constants 
for the figure-eight knot ($4_1$) and the trivial knot ($0$) 
versus the number of segments $N$. Fitting curve is given by 
$D_{4_1}/D_{0}=a(1+b N^{-c})$ where 
$a=1.02 \pm 0.56$, $b=1.76 \pm 8.26$, 
and $c=0.70 \pm 2.58$ with $\chi^2=0.03$. }
\end{figure}


\begin{figure}
  \begin{center}
    \begin{tabular}{cc}
    \resizebox{120mm}{!}{\includegraphics{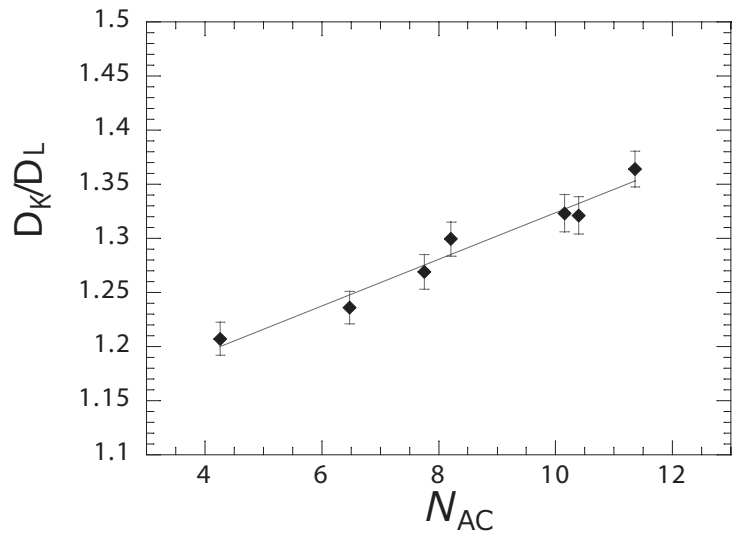}}
    \end{tabular}
    \label{fig:D_ACN}
  \end{center}
\caption{
$D_K/D_L$ versus 
the average crossing number ($N_{AC}$) of ideal knot $K$ for 
$N=45$: 
The data are approximated by 
$D_K/D_L = a + b N_{AC}$ where $a=1.11 \pm 0.02$ and 
$b=0.0215 \pm 0.0003$ with $\chi^2=2$. 
}
\end{figure}


\begin{figure}[htb]
  \begin{center}
    \begin{tabular}{cc}
  \resizebox{120mm}{!}{\includegraphics{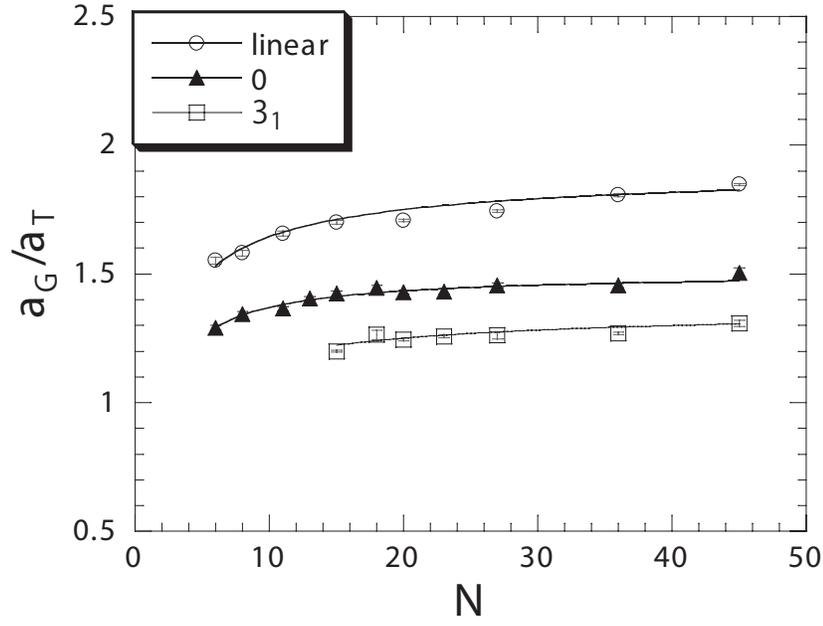}}
    \end{tabular}
    \label{fig:TG}
  \end{center}
\caption{$a_G/a_T$ of linear and knotted ring chains 
with knots $0$, $3_1$, $4_1$, $5_1$, $6_1$ and $7_1$, versus $N$. 
 Fitted by $a_G/a_T = a (1 - b N^{-c})$ with  
the following best estimates: For a linear chain, 
$a=2.37 \pm 0.29$, $b=0.52 \pm 0.03$,  
$c=0.22 \pm 0.10$, $\chi^2=55$; 
for the trivial knot ($0$),  
$a=1.51 \pm 0.03$, 
$b=0.72 \pm 0.23$, $c=0.89 \pm 0.22$, $\chi^2=14$; 
for the trefoil knot ($3_1$),  
$a=1.30 \pm 0.03$, 
$b=2.10 \pm 3.71$, 
$c=1.27 \pm 0.76$, $\chi^2=20$.}
\end{figure}

%
%

\end{document}